\documentclass{aa}
\usepackage{graphicx}

\begin{document}

\authorrunning{Lindborg et al.}

\titlerunning{Doppler images of II Peg during 1994-2002}

   \title{Doppler images of the RS CVn binary II Pegasi during the years 1994-2002 }

   \author{M. Lindborg \inst{1} \and 
           M. J. Korpi \inst{1} \and 
           T. Hackman \inst{1} \and
           I. Tuominen \inst{1} \and 
           I. Ilyin \inst{2} \and 
           N. Piskunov \inst{3}
	   }

   \offprints{M. Lindborg\\
          \email{marjaana.lindborg@helsinki.fi}
          }   

   \institute{Division of Geophysics and Astronomy, Department of Physics, PO 
Box 64 , FI-00014
              University of Helsinki, Finland
          \and Astrophysikalisches Institut Potsdam, An der Sternwarte
         16, 14882 Potsdam, Germany
         \and Department of Astronomy and Space Physics, Uppsala
         University, SE-751 20, Uppsala, Sweden }

\date{Received / Accepted}

\abstract{}
{We publish 16 Doppler imaging temperature maps for the years
  1994--2002 of the active RS CVn star II Peg. The six maps from
  1999-2002 are based on previously unpublished observations. Through
  Doppler imaging we want to study the spot evolution of the star and
  in particular compare this with previous results showing a
  cyclic spot behaviour and persistent active longitudes.}
{The observations were collected with the SOFIN spectrograph at the
  Nordic Optical Telescope. The temperature maps were calculated using
  a Doppler imaging code based on Tikhonov regularization.}
{During 1994-2001, our results show a consistent trend in the derived
  longitudes of the principal and secondary temperature minima over
  time such that the magnetic structure appears to  rotate
  somewhat more rapidly than the orbital period of this close binary.  A
  sudden phase jump of the active region occured
  between the observing seasons of 2001 and 2002.  No clear trend over
  time is detected in the derived latitudes of the spots, indicating
  that the systematic motion could be related to the drift of the spot
  generating mechanism rather than to differential rotation. The
  derived temperature maps are quite similar to the ones obtained
  earlier with a different methods, the main differences occurring in
  the spot latitudes and relative strength of the spot structures.}
{We observe both longitude and latitude shifts in the spot activity of 
II Peg. However, our results are not consistent with the periodic behaviour 
presented in previous studies. }

\keywords{stars: activity, imaging, starspots, HD 224085}
 
\maketitle

\section{Introduction}

Studying starspots helps us to better understand the magnetic
activity and related phenomena both in the Sun and  magnetically
more active stars. The two most common methods to study starspots are
Doppler imaging and time series analysis of photometry. With the
Doppler imaging method we can calculate surface temperature maps of
the star. The method is based on the fact that starspots on the
surface cause changes in photospheric absorption lines (Vogt \&
Penrod 1983; Piskunov 2008).

Suitable absorption lines for Doppler imaging of late-type stars are
neutral metal lines (e.g. Fe I, Ni I, Ca I) in the wavelength range
6000 \AA -- 7500 \AA. Important input parameters for Doppler imaging
are the projected rotational velocity, surface gravity, inclination,
microturbulence and orbital parameters.

II Pegasi (HD 224085) is one of the very active RS CVn stars. RS CVn stars
are closely detached binaries where the more massive
component is a G-K giant or subgiant and the secondary usually a subgiant
or dwarf of spectral class G to M (Huenemoerder et al. 2001).
Because of the low luminosity of the secondary, many RS CVn systems
appear as single-line binaries making their spectral analysis easier.
In close binaries the rapid rotation is
maintained by tidal forces due to the close companion.  Large
amplitude brightness variation of RS CVn stars imply the presence of
enormous starspots on their surfaces covering up to 50\% of the
visible disk. Dynamos of active stars appear to produce more high-latitude
spots and larger spots than inactive stars (Saar \& Baliunas 1992).
Also coronal X-ray and microwave emissions, strong
flares in the optical, UV, radio and X-ray are observed.

Cool spots on the stellar surface will locally alter the photospheric
absorption lines and continuum intensities. Through Doppler imaging
these rotationally modulated distortions can be inverted into
temperature maps (Vogt et al., 1987). Even though Doppler imaging is a
powerful method to study stellar surfaces in detail, the results
should always be interpreted critically. The method sufferes from
difficulties in modelling spectral lines in late-type stars. Any
errors in e.g. the stellar and spectral parameters will result in
systematic errrors in the results (Hackman et al. 2001).

II Pegasi has been spectroscopically monitored for almost 19 years
with the SOFIN spectrograph at NOT, La Palma, Spain. Previous studies of the
surface temperature distribution of II Peg include the papers of
Berdyugina et al. (1998a), who presented surface temperature maps for
1992-1996, and Bergyugina et al. (1999b) with surface maps for
1996-1999.  Both studies were based on observations with the
SOFIN spectrograph at NOT. Gu et al. (2003) presented surface images
of II Peg for 1999-2001 based on observations with the Coude \'echelle
spectrograph at the Xinglong station of the National Astronomical
Observatories, China, but the spectral lines used for inversions were
different from those of the SOFIN observations.

Photometric light curve variations of the star were analysed by
Berdyugina \& Tuominen (1998), and by Rodon\`o et al. (2000).  Berdyugina et 
al. (1998a,b, 1999a,b) and Berdyugina \& Tuominen (1998) 
present results that there were two active longitudes separated
approximately by 180$^{\circ}$, migrating in the orbital reference
frame, and that a switch of activity level occurs periodically with a
period of about 4.65 years. In the surface images of Gu et al. (2003) the
general spot pattern was quite similar, but the drift with respect of
the orbital frame was less obvious, and the switch of the activity
level appeared to occur earlier than predicted by Berdyugina et
al. Rodon\`o et al. (2000) found a much more complicated spot pattern
from their analysis of photometry: they report the existence of a
longitudinally uniformly distributed component together with three
active longitudes, with complicated cyclic behavior.

Spectropolarimetric observations in four Stokes parameters provide a
new basis for Doppler imaging techniques (cf. Semel 1989 and Piskunov 2008).  
Time
series of high resolution spectropolarimetric observations have
succesfully been used for retrieving maps of the surface magnetic
field for both late type stars, with temperature spots (Piskunov,
2008; Carroll et al. 2009) and Ap stars with chemical abundance spots
(Kochukhov et al. 2006).  Carroll et al. (2009) applied a
Zeeman-Doppler imaging technique to derive the magnetic field
configuration on the surface of II Peg during 2007 using
spectropolarimetric (Stokes I and V) observations with SOFIN. Their
maps show a very similar spot pattern as the one found by Berdyugina
et al. (1998b, 1999b); moreover, the radial field direction was
opposite on different active longitudes.

\section{Spectroscopic observations with SOFIN}

II Peg is part of our long-term observational projects. 
The time series of almost 19 years of observations of II Peg is
unique. For this study we used high resolution spectra of II Peg
measured in August 1994, July 1995, July 1996, October 1996, June
1997, August 1997, December 1997, July 1998, October 1998, November
1998, July-August 1999, September 1999, October 1999, August 2001,
August 2002 and November 2002.  All the observations were made using
the SOFIN high resolution \'echelle spectrograph at the 2.6 m Nordic
Optical Telescope (NOT), La Palma, Spain.  The data were acquired with
the 2nd camera equipped with a  EEV CCD detector of 1152$\times$770 pixels,
and fron January 2002 a new Loral one with 2048x2048 and the higher quantum
efficiensy.
The spectral region 6160-6210$\AA$ was chosen for Doppler imaging.
The observations are summarized in Table 1-3.

The spectral observations were reduced with the 4A software system
(Ilyin 2000). Bias, cosmic ray, flat field and scattered light
corrections, wavelength calibration and normalization and corrections for
the motion of the Earth were included in
the reduction process.

\begin{figure}
\begin{center}
\includegraphics[width=3.5in]{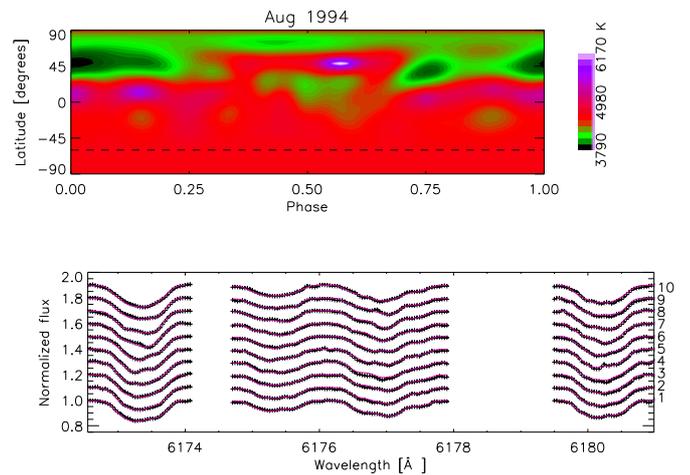}
\caption{Observing run during August 1994. Upper panel: Mercator
  projection of the obtained surface temperature distribution. Lower
  panel: observed (crosses) and calculated line profiles (solid
  lines).}\label{Aug1994}
\end{center}
\end{figure}

\begin{figure}
\begin{center}
\includegraphics[width=3.5in]{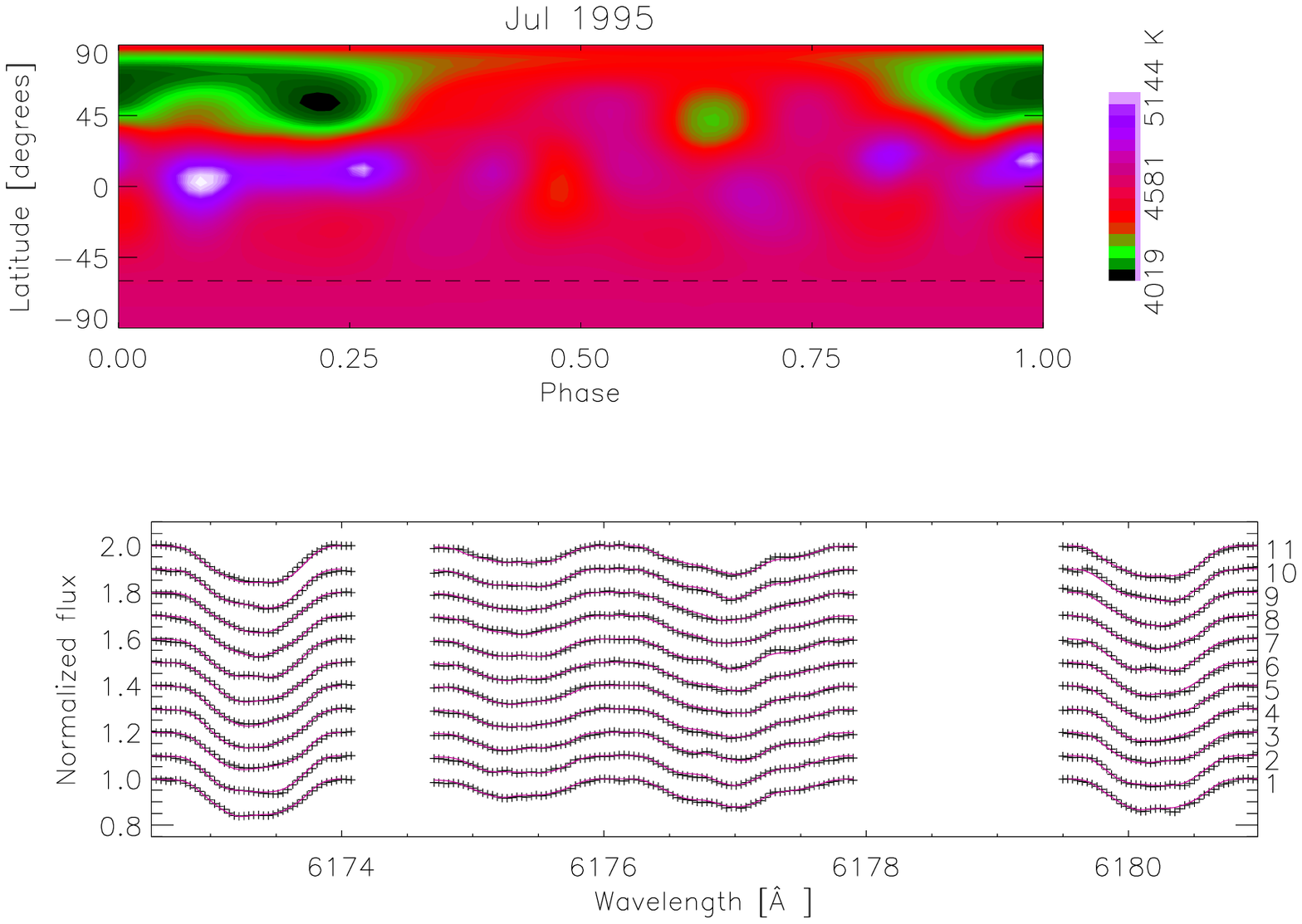}
\caption{Observing run during July 1995.}\label{Jul1995}
\end{center}
\end{figure}

\begin{figure}
\begin{center}
\includegraphics[width=3.5in]{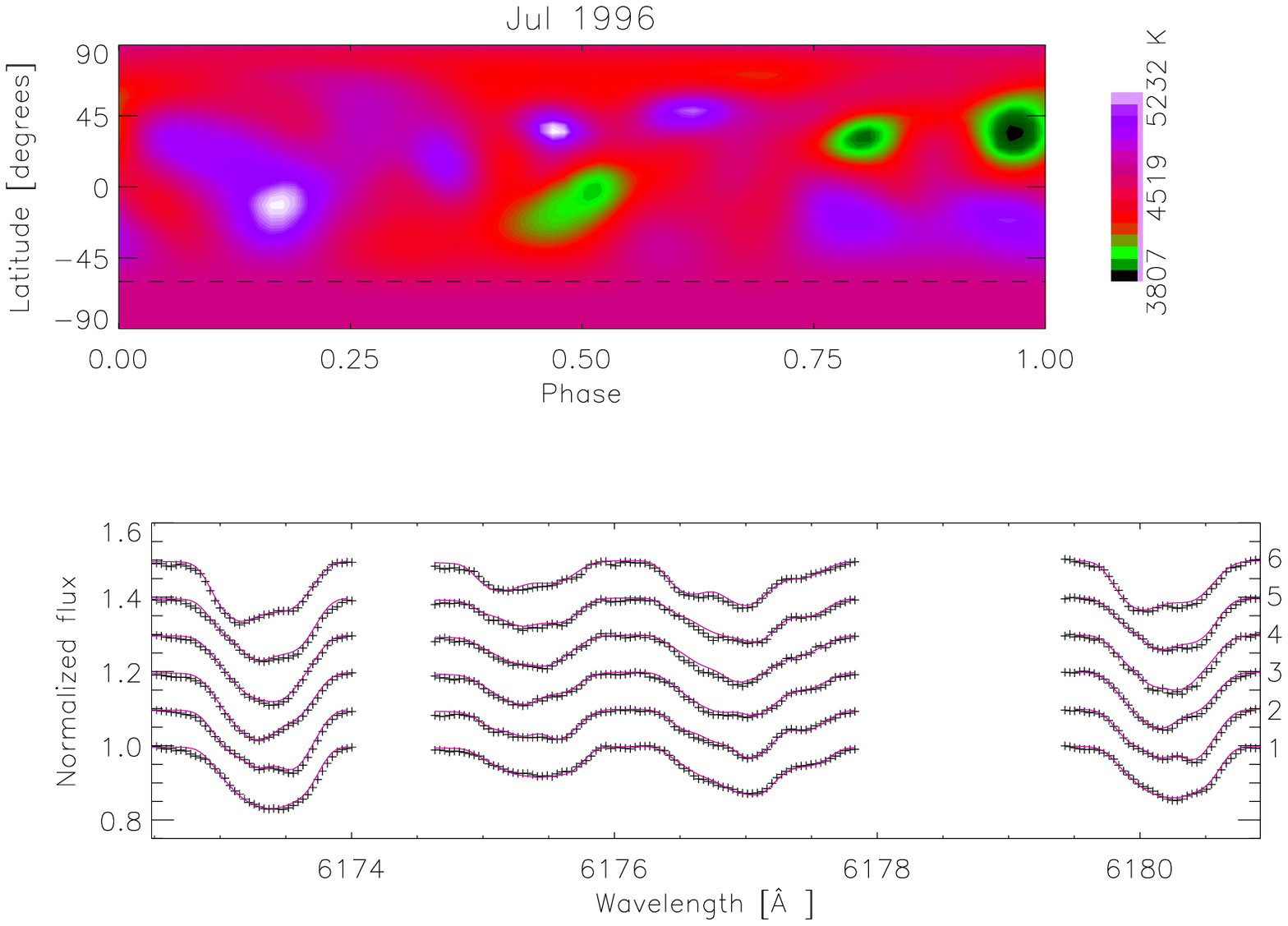}
\caption{Observing run during July 1996.}\label{Jul1996}
\end{center}
\end{figure}

\begin{figure}
\begin{center}
\includegraphics[width=3.5in]{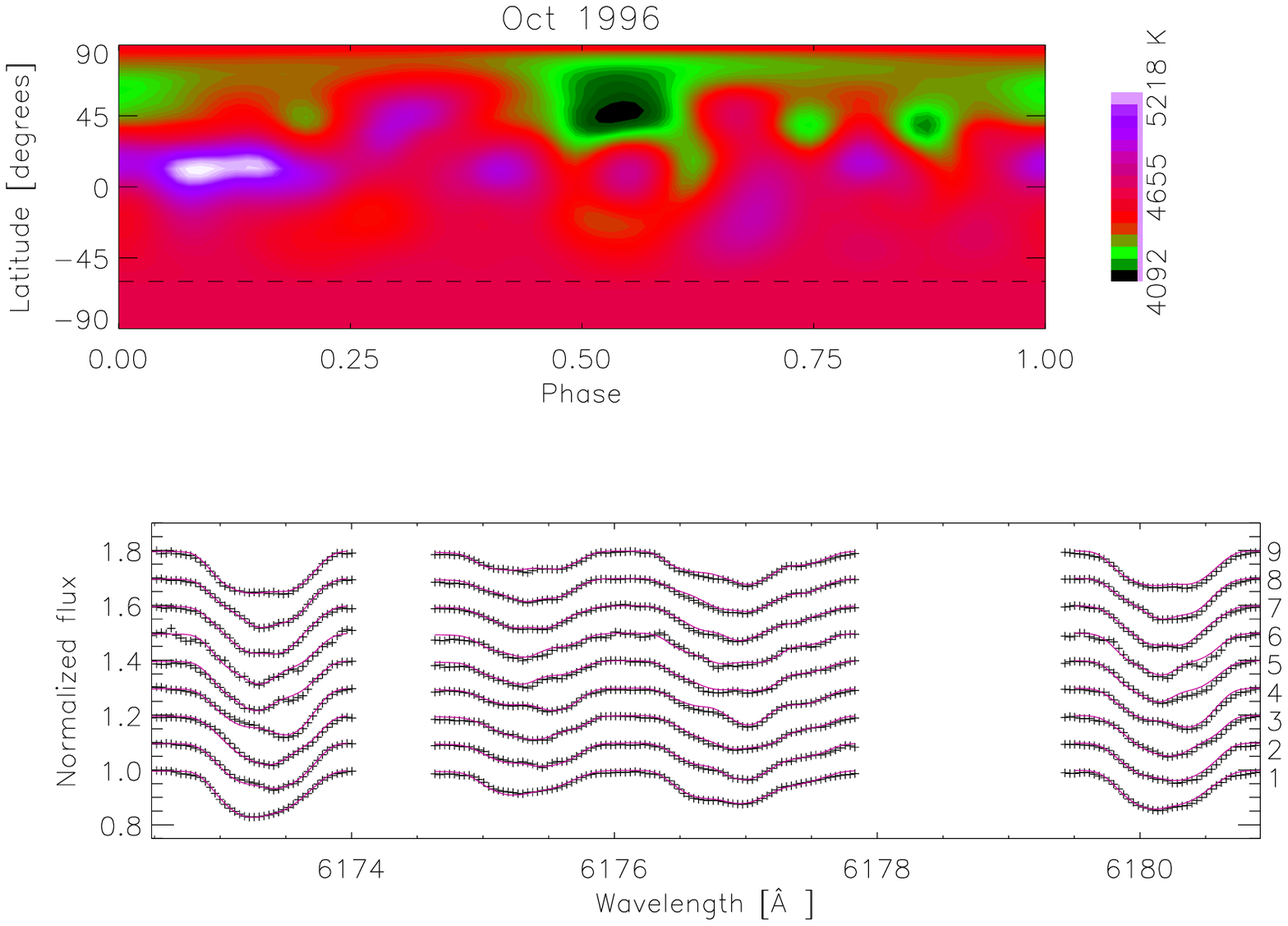}
\caption{Observing run during October 1996.}\label{Oct1996}
\end{center}
\end{figure}

\begin{figure}
\begin{center}
\includegraphics[width=3.5in]{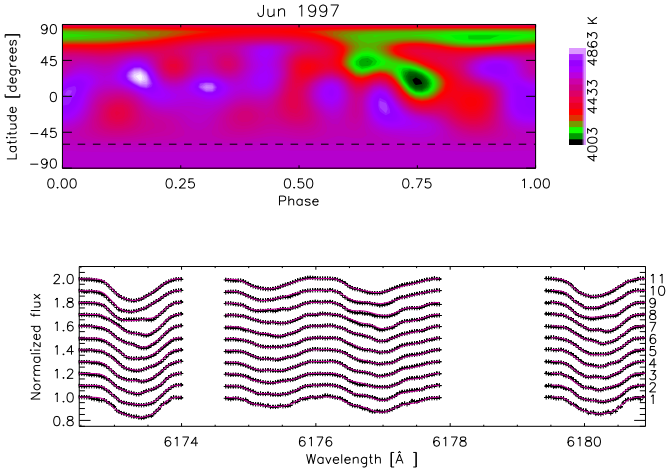}
\caption{Observing run during June 1997.}\label{Jun1997}
\end{center}
\end{figure}

\begin{figure}
\begin{center}
\includegraphics[width=3.5in]{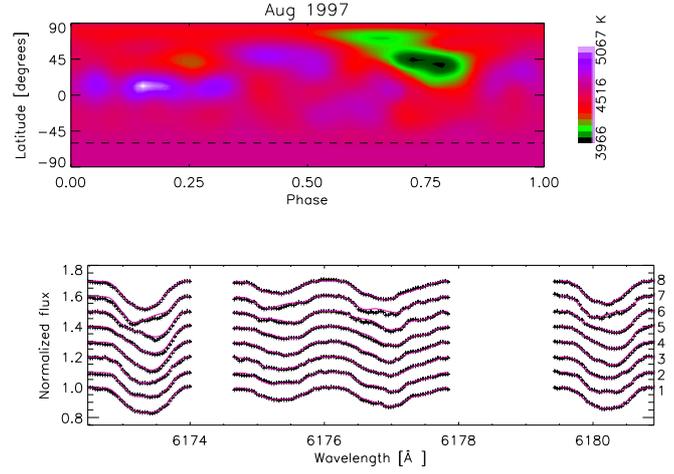}
\caption{Observing run during July-August 1997.}\label{Aug1997}
\end{center}
\end{figure}

\begin{figure}
\begin{center}
\includegraphics[width=3.5in]{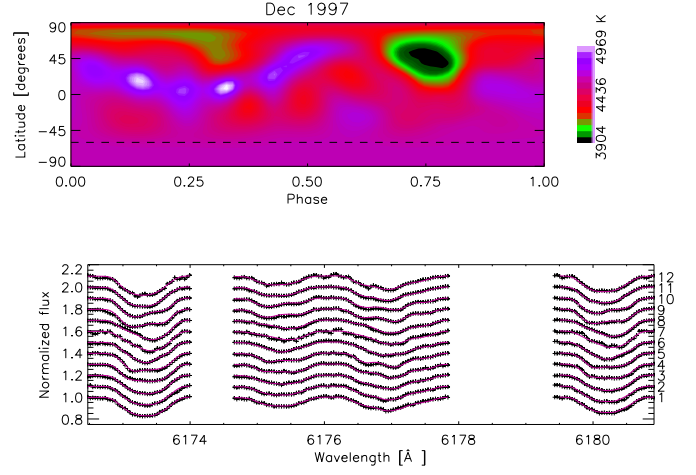}
\caption{Observing run during December 1997.}\label{Dec1997}
\end{center}
\end{figure}

\begin{figure}
\begin{center}
\includegraphics[width=3.5in]{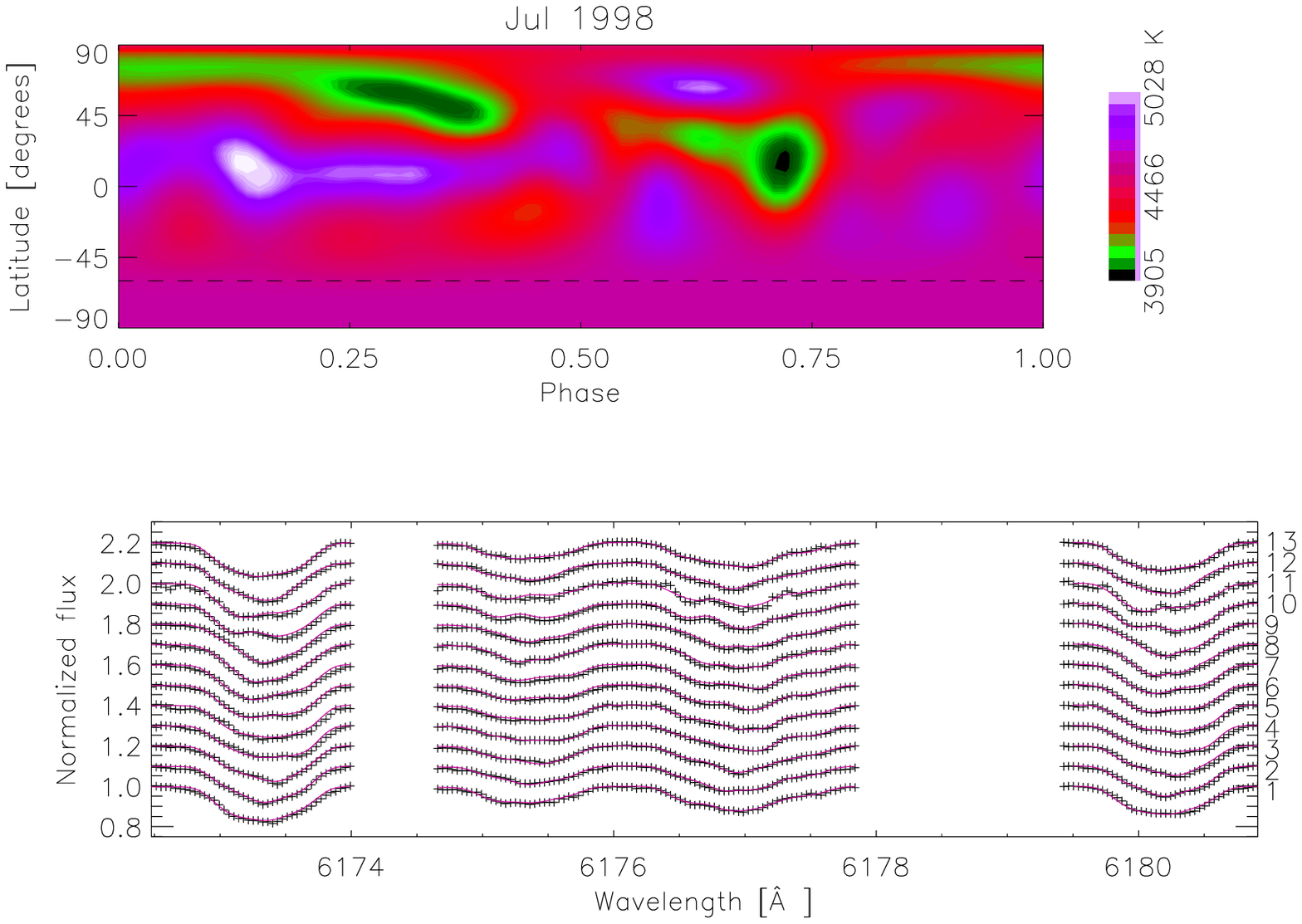}
\caption{Observing run during July 1998.}\label{Jul1998}
\end{center}
\end{figure}

\begin{figure}
\begin{center}
\includegraphics[width=3.5in]{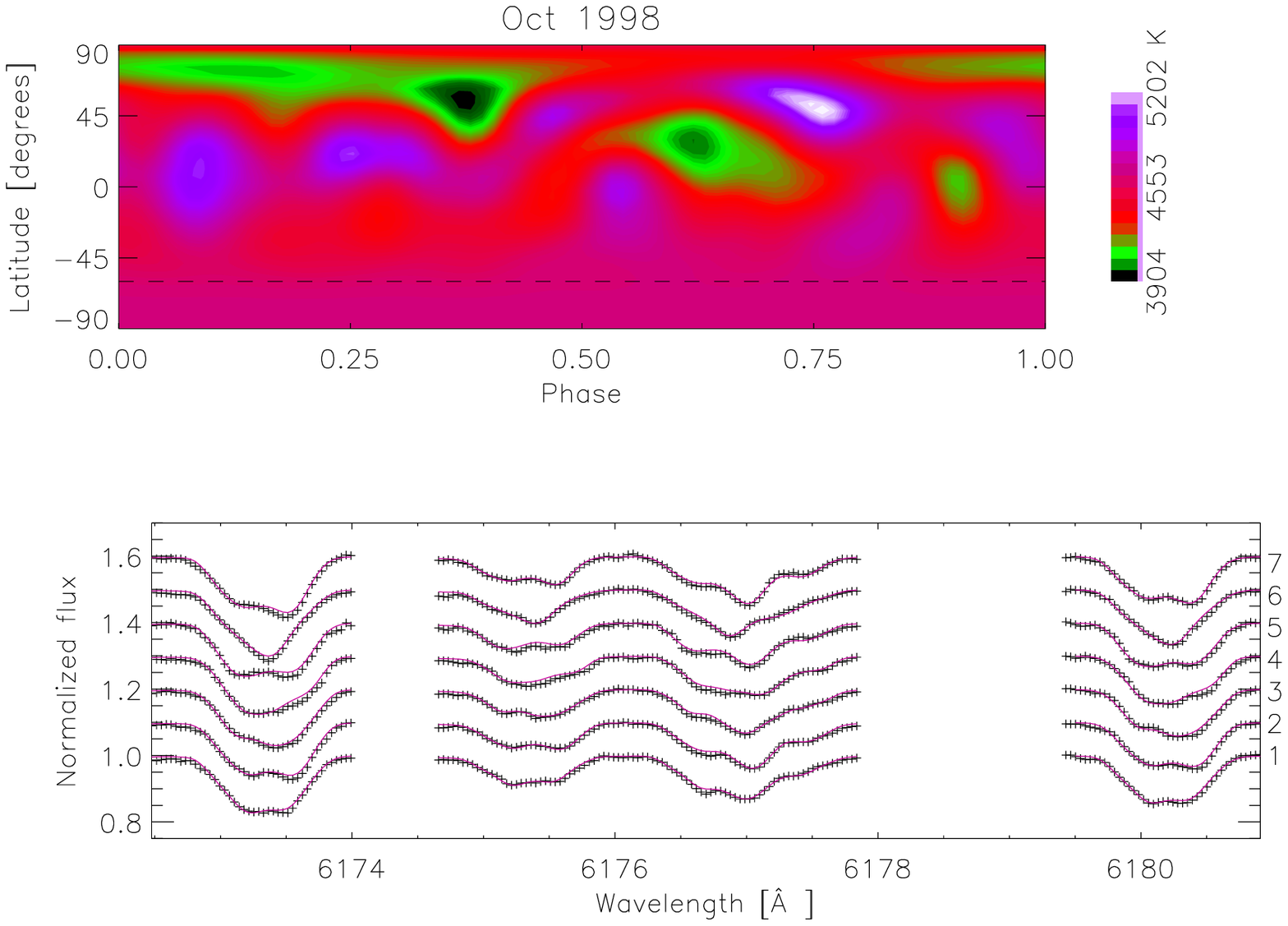}
\caption{Observing run during October 1998.}\label{Oct1998}
\end{center}
\end{figure}

\begin{figure}
\begin{center}
\includegraphics[width=3.5in]{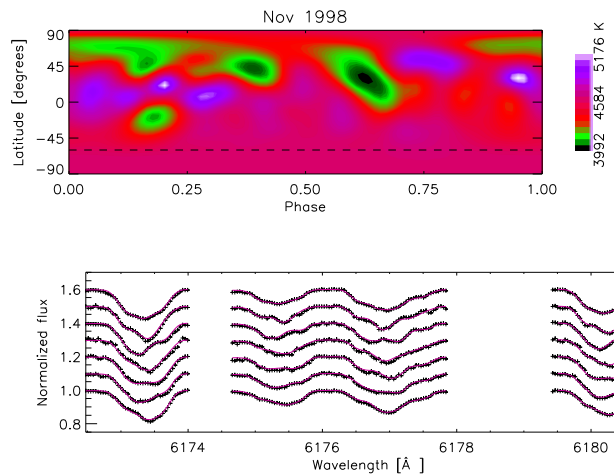}
\caption{Observing run during November 1998.}\label{Nov1998}
\end{center}
\end{figure}

\begin{figure}
\begin{center}
\includegraphics[width=3.5in]{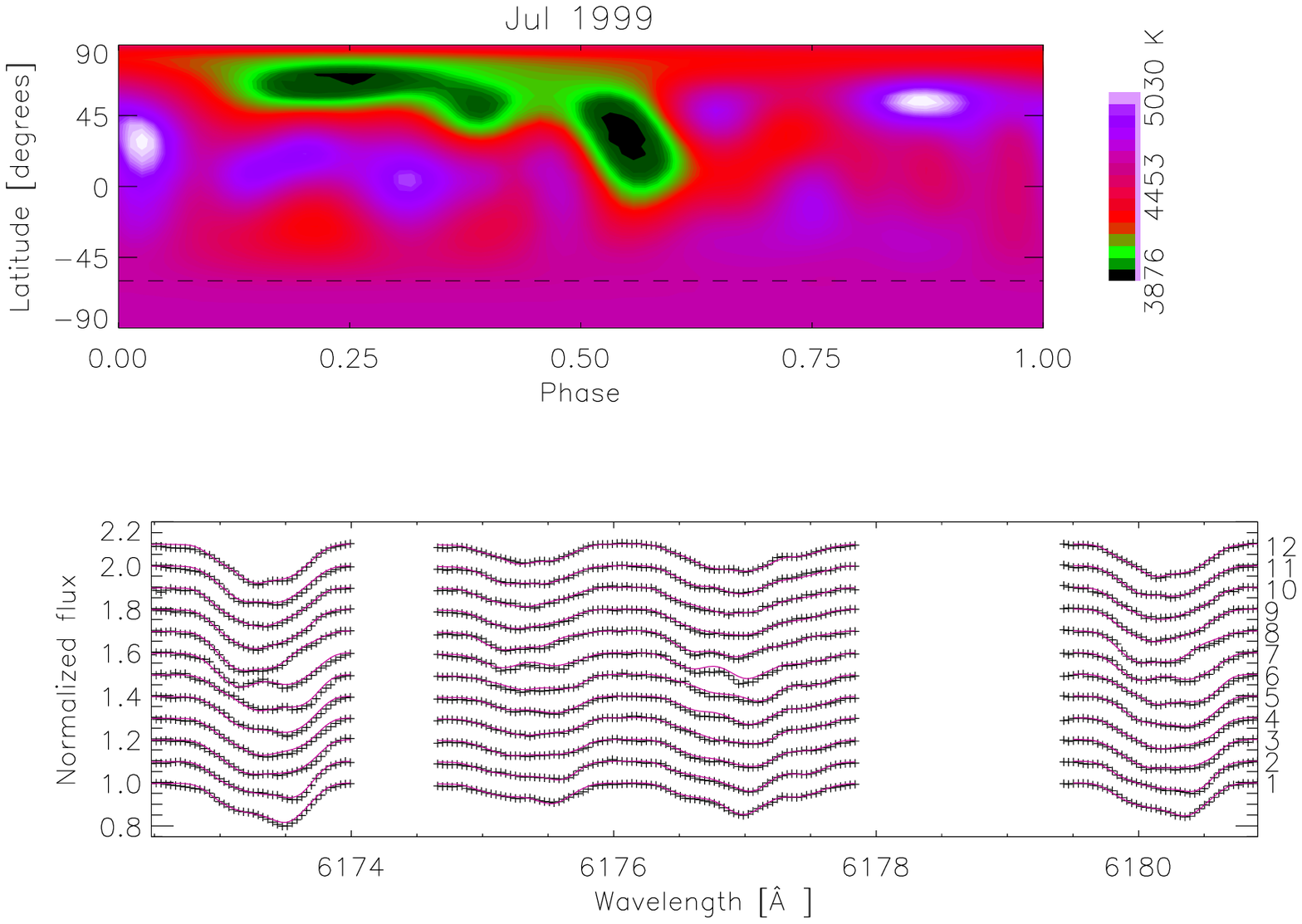}
\caption{Observing run during July 1999.}\label{Jul1999}
\end{center}
\end{figure}

\begin{figure}
\begin{center}
\includegraphics[width=3.5in]{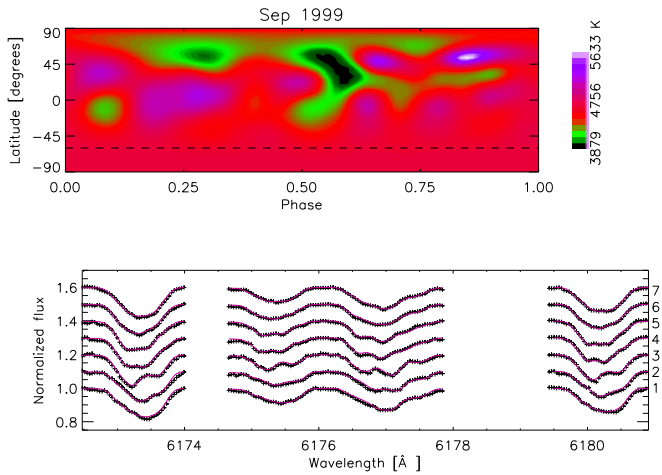}
\caption{Observing run during September 1999.}\label{Sep1999}
\end{center}
\end{figure}

\begin{figure}
\begin{center}
\includegraphics[width=3.5in]{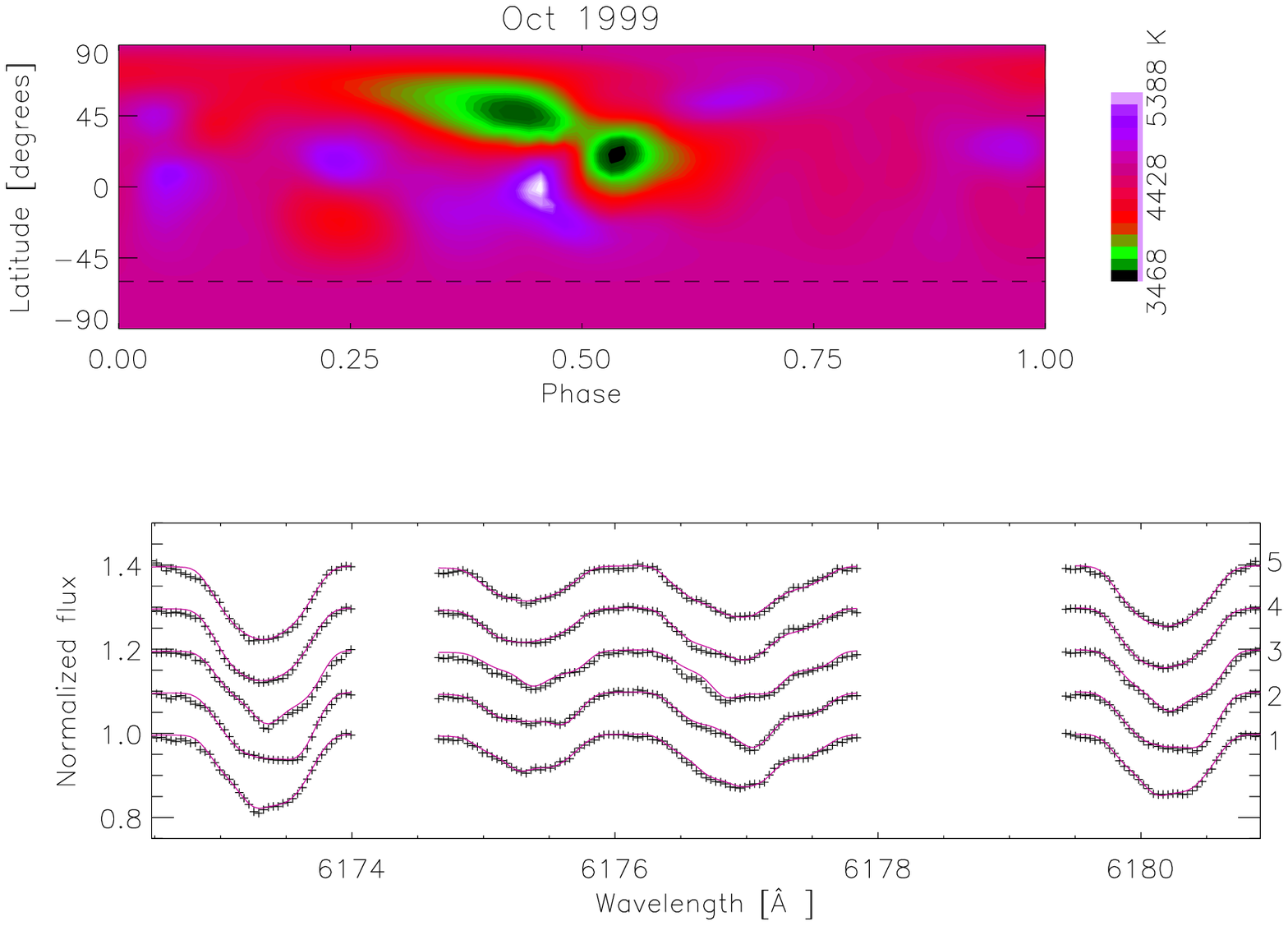}
\caption{Observing run during October 1999.}\label{Oct1999}
\end{center}
\end{figure}

\begin{figure}
\begin{center}
\includegraphics[width=3.5in]{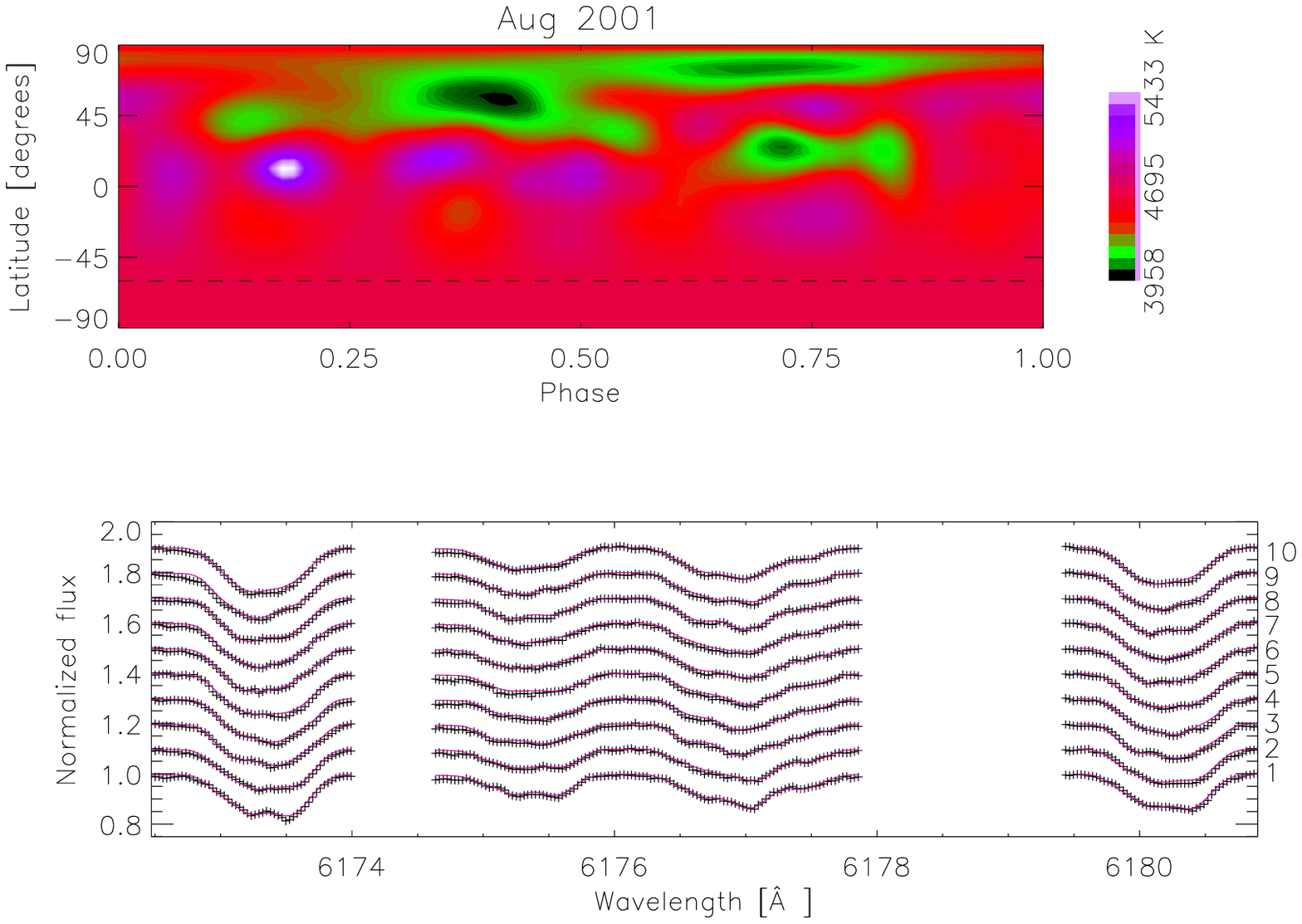}
\caption{Observing run during August 2001.}\label{Aug2001}
\end{center}
\end{figure}

\begin{figure}
\begin{center}
\includegraphics[width=3.5in]{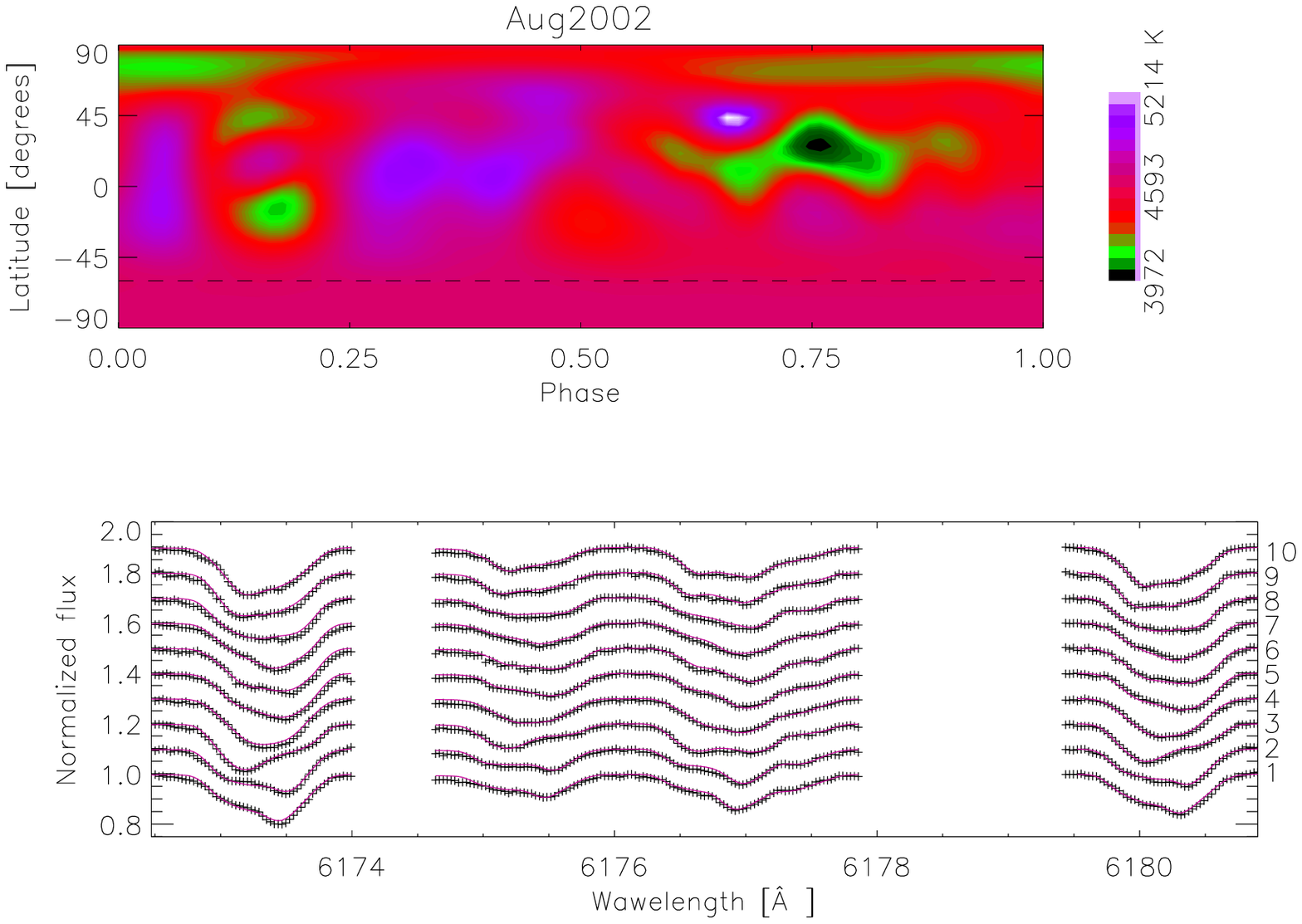}
\caption{Observing run during August 2002.}\label{Aug2002}
\end{center}
\end{figure}

\begin{figure}
\begin{center}
\includegraphics[width=3.5in]{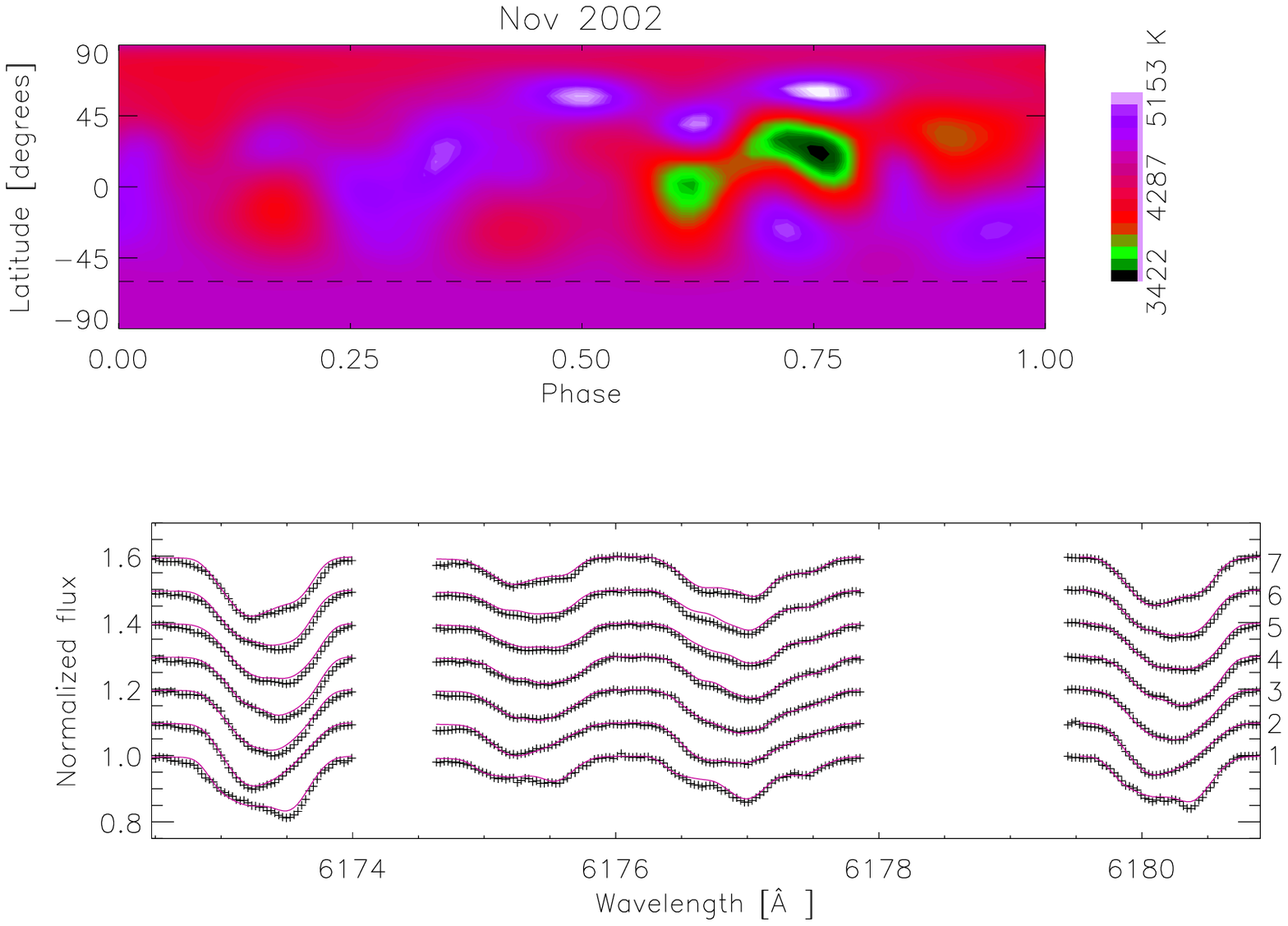}
\caption{Observing run during November 2002.}\label{Nov2002}
\end{center}
\end{figure}

\section{The Doppler imaging technique}

Doppler imaging utilizes the fact that
under rapid rotation surface spots will induce bumps in the
rotationally broadened absorption line profiles. As the star rotates,
these bumps will move across the line profiles. Using a set of
observations distributed over the rotational phases we can solve the
inverse problem and search for such a temperature distribution that
reproduces the observations (Vogt et al. 1987).

We use the Doppler imaging technique developed by Piskunov (the code
{\sc INVERS7}, Piskunov et al. 1990; Piskunov 1991). In this method
the otherwise ill-posed inversion problem is solved by Tikhonov
regularization. In practice this means introducing an additional
structure constraint minimizing the surface gradient of the solution.
The Doppler imaging solution is retrieved using a table of calculated local 
line profiles for different surface positions (limb angles) and different 
effective temperatures. The local line profiles are calculated using numerical
models of stellar atmospheres. In the analysis presented here,
we used the stellar model atmospheres of Kurucz (1993)
with temperatures 3000 -- 6000 K and gravity $\log g = 4.0$.

The stellar parameters used in the inversions were chosen mostly
according to Berdyugina et al. (1998b), and read $\xi_t=2.2$
kms$^{-1}$, $\zeta_t$=3.5 kms$^{-1}$, $P_{\rm orb}$=6.724333 d, $v
\sin i$=22.6 kms$^{-1}$, $i=60^{\circ}$.

The following spectral lines were used in the surface temperature
inversions: Fe I 6173.0050 \AA, Ni I 6175.3600 \AA, Ni I 6176.8070 \AA,
Ni I 6177.2560 \AA, and  Fe I 6180.2030 \AA. 
All these lines are actually blends of lines. All in all we
included 
30 lines within the wavelength range 6172--6182 {\AA} in the analysis.
 Spectral
line parameters were obtained from the Vienna Atomic Line Database
(Kupka et al. 1999).
Solar element abundance values and standard spectral parameters were used as a
starting point. A preliminary analysis showed, that in particular the
Ni abundance had to be adjusted from the solar value of $-5.79$ to $-5.50$.
The $\log(gf)$ values of two included Ni I
lines were modified from the standard values of $-0.53$ 
to 
$-0.540$ (Ni I 6175.360 {\AA}) and  $-0.390$ (Ni I 6176.807 {\AA}) as the 
standard values were found to produce a wrong balance between these lines.

\section{Results}

\begin{figure*}
\begin{center}
\includegraphics[width=3.5in]{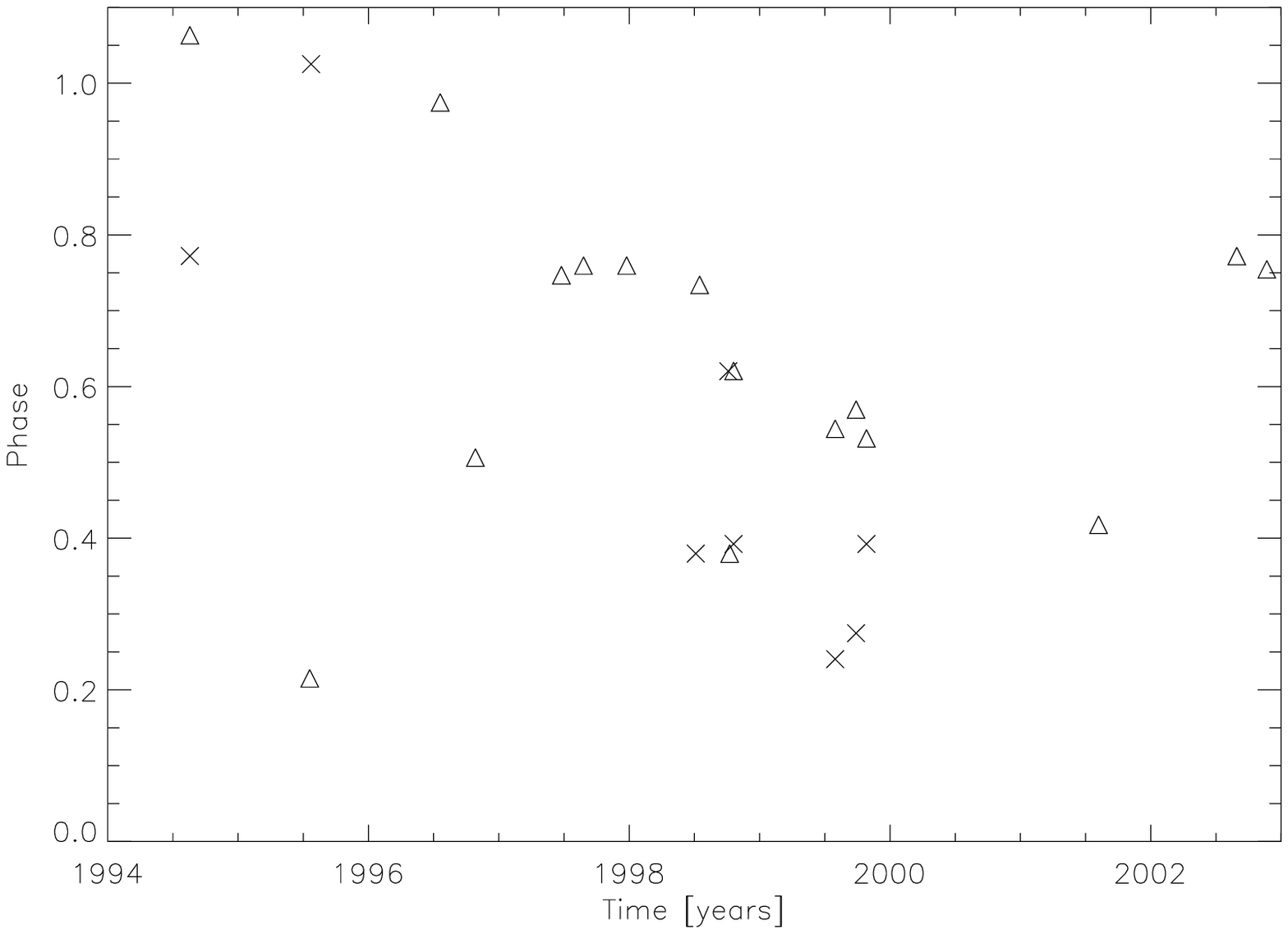}\includegraphics[width=3.5in]{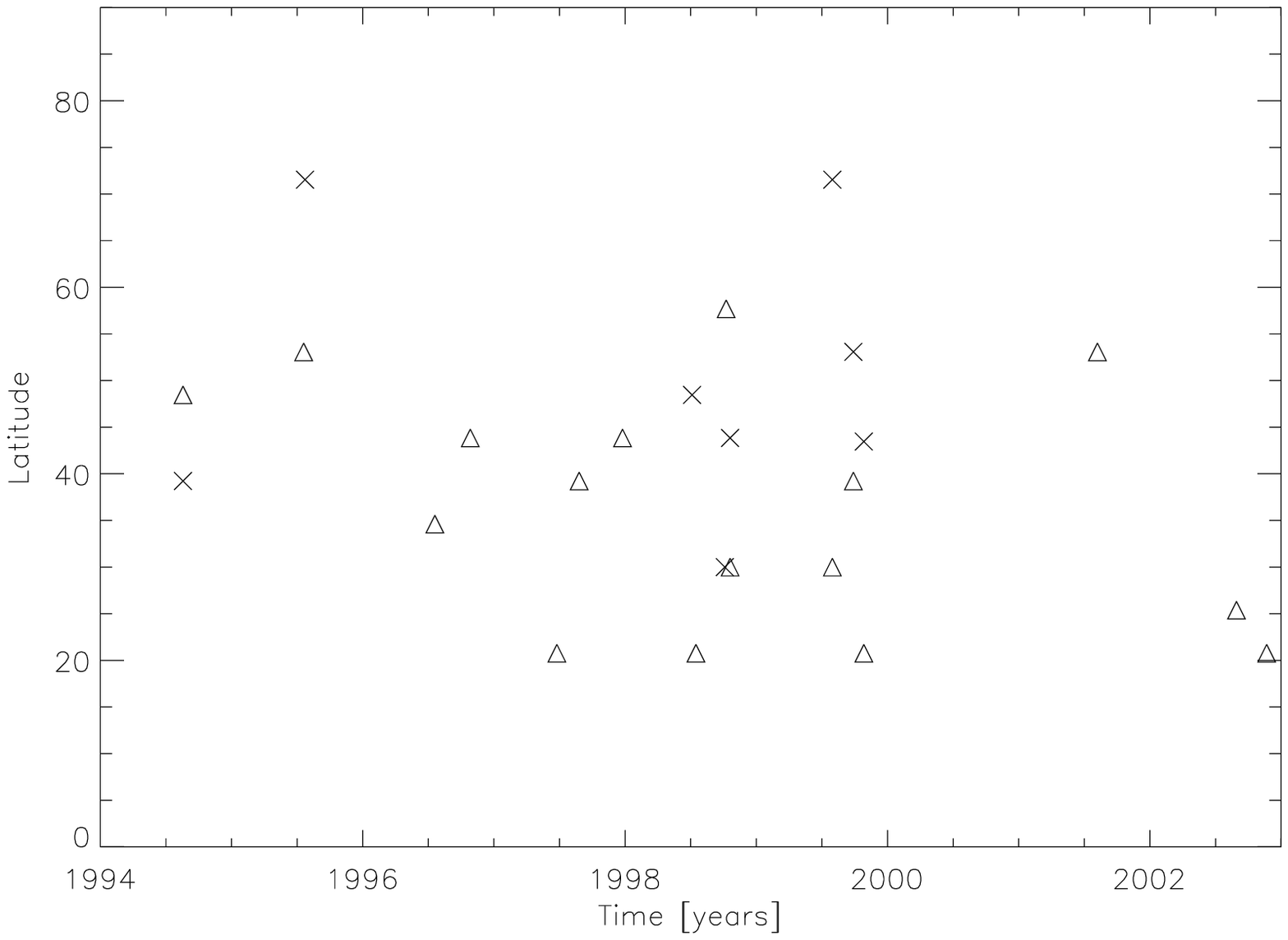}
\caption{The longitude (left panel) and latitude (right panel) of the primary (triangles) and
  secondary (crosses) minima determined from the surface temperature
  maps as function of time.}\label{spotpos}
\end{center}
\end{figure*}

Figures \ref{Aug1994}-\ref{spotpos} show the results of the
inversions; in Figs. \ref{Aug1994}-\ref{Nov2002}, we show the derived
temperature maps, and in Fig. \ref{spotpos}, the derived longitudes
(actually rotation phases, left panel) and latitudes (right panel) versus time.

During August 1994 (Fig.~\ref{Aug1994}) an extended spot structure is
visible, located at mid-latitudes. The primary minimum of the
temperature occurs nearly at phase 0.06, and a secondary minimum
roughly at the phase 0.77. Our map is in reasonable agreement with the
one of Berdyugina et al. (1998a) from the same observations. Roughly 11
months later (July 1995, Fig.~\ref{Jul1995}), the extended structure
is still visible, although located at somewhat higher latitudes and apparently 
shifted by $\sim$ 0.2 in phase. The
temperature minima are not so pronounced as in 1994, and the deeper
minimum is now located roughly at 0.22, and the secondary one roughly
at the position of the former primary minimum at 0.025. Again, the
main features are similar to the maps derived by Berdyugina et
al. (1998a); this is to be expected, as identical observations have
been used.

During the observing run of July 1996 (Fig.~\ref{Jul1996}), the phase
coverage is rather poor. The extended structure that persisted during
the 1994 and 1995 observations is no longer visible; the cool regions are
now seen in the form of smaller spots at relatively low latitudes. The
primary minimum of temperature occurs approximately at the phase 0.97
and latitude 35 degrees. The other two spots located at phases 0.78
and 0.5 are less significant, and therefore we leave them out of
Fig.~\ref{spotpos}. A few months later (October 1996,
Fig.~\ref{Oct1996}), cool regions can be seen at approximately same
locations. Their relative strength, however, has dramatically changed:
the former deepest minimum now appears as almost insignificant, while
the formerly weakest structure at 0.5 is now clearly the strongest and
largest one at 0.51. The strongest minimum appears at somewhat higher
latitude than in July; the time between the observing runs is short,
due to which the difference in the principal temperature minimum
longitude cannot be explained by differential rotation due to changing
latitude. A more likely explanation is related to short time-scale
changes of the activity level of the spots.

Our maps for July and October 1996 are not in a
very good agreement with the ones obtained by Berdyugina et
al. (1998a): their maps showed two spots, the principal one being located
within phases 0.8-1.0 for both July and October 1996.  Concerning the
maps of July 1996, Berdyugina et al. had additional data from the
Crimean Coud\'e spectrograph, so their phase coverage was better. For
October 1996, the data sets were, however, identical, except one
observation with very poor signal to noise ratio included in our
inversion but excluded in theirs. 

During the observing seasons of 1997 (June, Fig.~\ref{Jun1997};
August, Fig.~\ref{Aug1997}; December, Fig.~\ref{Dec1997}), the
activity is concentrated around the phase 0.75 in all the three
maps. During June 1997, the spot structure is more elongated towards
lower phases, the principal minimum being at a fairly low latitude of 21
degrees. In August 1997 the structure is still similarly elongated,
but a general trend is that the whole structure is located somewhat
higher in latitude (39 degrees). In December 1997, the lowest
temperature is still around the same phase, but it has moved towards
higher latitudes (44 degrees). The principal temperature minimum we
have recovered matches well the one obtained by Berdyugina et
al. (1999b) from identical observations (except for the phase 0.92
missing from our analysis during June 1997). In their maps, however, a
clear secondary minimum can be observed around the phase 0.25, almost
entirely absent in our maps. The spots are also consistently at higher
latitudes in Berdyugina et al. maps than in ours.

In July 1998 (Fig.~\ref{Jul1998}) the primary spot structure, half a
year earlier located at phase 0.75, has clearly decreased in magnitude,
while the weaker structure, now at phase 0.38, has grown stronger, the
structures being nearly of the same strength in July 1998. The
principal minimum has rather a low latitude of 21 degrees, while the
secondary minimum occurs higher, at the latitude 48. A similar
configuration is seen in the map for October 1998
(Fig.~\ref{Oct1998}), although the structures have moved closer to
each other in longitude (phases 0.38 and 0.62), and swapped their
strength relative to each other. In November 1998
(Fig.~\ref{Nov1998}), however, the spot structure at the phase 0.62 is
again stronger - in our maps, the relative strength of the spot
structures is much more variable than in the maps of Berdyugina et
al. (1999b), who report only one change of the relative strengths
having occurred between the observing seasons of Dec 1997 and Jul
1998, (flip-flop phenomenon). We also note that the phase coverage for
the observing seasons is poor for October and November 1998. 

Signs of peculiar apparently bright features, possibly related to outbursts 
such as stellar flares, can be seen as ``absorption dips'' in the spectra at 
phase 0.78 in October (label 6 in Fig.~\ref{Oct1998}) and phase 0.55 in 
November 1998 (label 4 in Fig.~\ref{Nov1998}. Therefore, these maps may
be regarded as less reliable than for July 1998. Flares on II Peg
in the optical spectral lines have been reported several times
(Berdyugina et al., 1999a) and the surface patterns might change,
including positions of the active regions, during optical flare
events.

During the year 1999 (July-August 1999 Fig.~\ref{Jul1999}, September
1999 Fig.~\ref{Sep1999} and October Fig.~\ref{Oct1999}) an extended
spot structure is seen in all maps, roughly in between the phases 0.2
and 0.6. The minimum temperature occurs around the phase 0.55 in all
three maps, and its latitude is relatively low. A secondary wing-like
structure is also visible, located at higher latitudes (72 degrees)
and at the phase 0.24 during July 1999. In September, this structure
has moved towards lower latitudes (53 degrees) and higher phase
(0.27), and in October, even further in the same directions (latitude
of 43 degrees and phase 0.39). We note that the phase coverage of the
September and October observing runs is quite poor, so these maps are
not as reliable as the July-August 1999 map.

According to the flip-flop phenomenon presented in Berdyugina et
al. (1999b), the highest activity should have occured near the phase
zero during the year 1999 (see the dashed line approximating the path
of the stronger active longitude in their Fig. 6). We find no spots
near that phase. Moreover, the deepest temperature minimum occurs near
the
secondary minimum branch of Berdyugina et al. (1999b). Therefore, no
clear support for the flip-flop phenomenon with the spot cycle
proposed by Berdyugina et al. (1999b) can be found from our maps.

Gu et al. (2003) have also published temperature maps for the same
epoch, from a different observational facility and Doppler imaging
method. During July and August 1999, they found an extended
high-latitude structure between the phases (0.35 and 0.75), the
primary minimum being located near the phase 0.4, the secondary one at
roughly the phase 0.75. The location of the primary minimum matches
well the one derived from our maps; the position of the secondary one,
however, is different by 180 degrees.

Roughly 22 months later, in August 2001, (Fig.~\ref{Aug2001}) the
primary spot is still quite close to the phase which it had during the
1999 observing seasons (moved from the phase of 0.55 to 0.42). The
latitude of the primary spot is somewhat higher in 2001, namely 53
degrees. One year later, in August 2002, the deepest temperature
minimum has moved to the phase 0.75, where it seems to persist also in
the map from the November 2002 observations. This appears as a sudden jump
in the left panel of Fig.~\ref{spotpos} in an otherwise continuous
trend in the spot locations more or less consistent with a stable spot
structure moving in the orbital frame with a slightly shorter period
(more rapidly).  

In the right panel of Fig.~\ref{spotpos} we plot the latitudes of the
spot structures as a function of time. No systematic trends are visible
in this plot, indicating that the systematic change of the spot
positions during 1994-2001 cannot be attributed to differential
rotation of spots at different latitudes. The spots seen in the maps
of 2002, however, show clearly lower latitudes than during 2001.  For
a solar-type differential rotation pattern (equator rotates more rapidly
than the pole), such a change in latitude would result in a speed up
in the rotation of the spot structure; now the spot structure seems to
be slowing down with respect to the systematic trend. Therefore, at
least for solar-type differential rotation laws, the change in
latitude is not a very likely cause of the sudden change in the
longitudinal position.

The maps of Gu et al. (2003) for February 2000 and November 2001
reveal a systematic drift of the primary spot location towards higher
phases. Although the maps are not exactly of the same time epochs,
their trend seems to be consistent with the one we see.

\section{Conclusions}

In the present paper we give 16 surface images of II Pegasi for the
years 1994-2002, the 6 latest ones being based on yet unpublished
SOFIN@NOT observations. The earlier observations, analysed with a
different Doppler imaging method, have already been published by
Berdyugina et al. (1998a, 1999b). 

All in all our maps show quite good resemblance to the ones obtained
by Berdyugina et al. (1998a) and (1999b) with a different method.
The main differences between our results to the ones of Berdyugina et
al. are the consistently lower spot latitudes in our results. Some differences 
are also seen in the relative strength of
the spot structures, even during epochs when almost identical
observational data was used. Partly due to this, no definite signs of
a flip-flop event in between December 1997 and July 1998 can be
seen. 

In contrast, the derived spot positions from our maps are consistent
with a systematic drift of the spot structures in the orbital frame
during (1994-2001). The spot generating structure seems to be rotating
somewhat more rapidly than the binary system. Between the observing seasons
of 2001 and 2002, a sudden jump of spot locations towards larger
phases has occured. This is more consistent with the trend seen by Gu
et al. (2003) from a different set of observations for 1999-2001. As
no systematic change in the spot latitudes over time can be seen, we
cannot attribute this jump being related to differential rotation. Some
fast changes might be a consequence of a sudden outburst like a flare.

\begin{table*}
    \caption{Summary of the observations 1994 - 1995.}
    \label{table1}
    \begin{tabular}{lccclccc}

%
    \hline
    HJD      &Phase &S/N  &Label &HJD &Phase &S/N &Label\\
    2440000+ &      &     &      &2450000+ &&& \\ \hline
    August 1994 &&& &July 1995 &&& \\ \hline
    9578.6835   &0.3690 &286 &3  &9910.6629 &0.7388 &386 &8 \\
    9579.6544   &0.5134 &273 &5  &9911.6449 &0.8849 &357 &10 \\
    9580.6080   &0.6552 &285 &7  &9912.6448 &0.0336 &299 &1 \\
    9581.5425   &0.7941 &321 &9  &9913.6311 &0.1802 &293 &2 \\
    9582.5406   &0.9426 &251 &10 &9914.6525 &0.3321 &332 &4 \\
    9583.6266   &0.1041 &353 &1  &9915.6549 &0.4812 &295 &6 \\
    9584.6549   &0.2570 &371 &2  &9916.6640 &0.6313 &273 &7 \\
    9585.6606   &0.4066 &391 &4  &9917.6137 &0.7725 &251 &9 \\
    9586.6594   &0.5551 &425 &6  &9918.6255 &0.9230 &273 &11 \\
    9587.6787   &0.7067 &293 &8  &9920.6676 &0.2267 &320 &3 \\
                &       &    &   &9921.6362 &0.3707 &258 &5 \\\hline

    \end{tabular}
\end{table*}

\begin{table*}
  \begin{center}
    \caption{Summary of the observations 1996-1998.}
    \label{table2}
    \begin{tabular}{lccclccc}
    \hline
    HJD      &Phase &S/N  &Label &HJD &Phase &S/N &Label\\
    2450000+ &      &     &      &2450000+ &&& \\ \hline
    July 1996   &&& &October 1996 &&& \\ \hline
    293.6544 &0.6949 &273 &4 &381.3800 &0.7409 &242 &7\\
    294.6784 &0.8472 &198 &5 &382.4175 &0.8952 &264 &9\\
    295.6166 &0.9867 &228 &6 &383.4008 &0.0414 &282 &1\\
    297.6716 &0.2923 &253 &1 &384.3821 &0.1873 &274 &2\\
    298.6767 &0.4418 &302 &2 &385.4396 &0.3446 &259 &3\\
    299.6952 &0.5932 &253 &3 &386.4171 &0.4900 &275 &4\\
             &       &    &  &387.4776 &0.6477 &163 &5\\
             &       &    &  &387.5683 &0.6612 &94  &6\\
             &       &    &  &388.3760 &0.7813 &286 &8 \\ \hline
    June 1997 &&& &August 1997 &&&\\ \hline
    617.6833 &0.8822 &225 &10  &676.7269 &0.6628 &249 &5 \\
    618.7239 &0.0369 &153 &1   &677.7541 &0.8156 &112 &7 \\
    619.6745 &0.1784 &216 &3   &678.7250 &0.9599 &191 &8 \\
    620.6783 &0.3276 &307 &5   &679.7443 &0.1116 &234 &1 \\
    621.6597 &0.4736 &293 &7   &680.7498 &0.2611 &179 &2 \\
    622.7085 &0.6296 &285 &8   &681.7510 &0.4099 &160 &3 \\
    623.7143 &0.7791 &236 &9   &682.7517 &0.5588 &224 &4 \\
    624.6985 &0.9255 &329 &11  &683.7536 &0.7078 &175 &6 \\
    625.6800 &0.0715 &277 &2   &&&& \\
    626.7049 &0.2239 &298 &4   &&&& \\
    627.6788 &0.3687 &301 &6   &&&& \\ \hline
    December 1997 &&&&&&& \\ \hline 
    794.3720 &0.1584 &249 &3&&& \\
    795.3692 &0.3067 &218 &4&&& \\
    796.3834 &0.4575 &179 &6&&& \\
    802.4216 &0.3555 &208 &5&&& \\
    804.3410 &0.6409 & 66 &7&&& \\
    804.4534 &0.6577 &158 &8&&& \\
    805.3263 &0.7875 &170 &9&&& \\
    805.4735 &0.8094 &158 &10&&& \\
    806.3601 &0.9412 &164 &11&&& \\
    806.4636 &0.9566 & 98 &12&&& \\
    807.3734 &0.0919 &147 &1&&& \\
    807.4011 &0.0960 &129 &2&&& \\ \hline
    July 1998 &&& &October 1998 &&& \\\hline
    997.7161  &0.3982 &234 &6  &1088.4366 &0.8898 &244 &7 \\
    998.7255  &0.5483 &196 &8  &1089.4187 &0.0358 &258 &1 \\
    999.7027  &0.6937 &245 &10 &1090.4712 &0.1923 &277 &2 \\ 
    1000.6999 &0.8420 &208 &12 &1091.4116 &0.3322 &263 &3 \\
    1001.6995 &0.9907 &259 &13 &1092.4301 &0.4837 &190 &4 \\ 
    1002.7031 &0.1399 &230 &2  &1093.4439 &0.6344 &206 &5 \\
    1003.6983 &0.2879 &245 &4  &1094.4554 &0.7848 &240 &6 \\
    1004.7087 &0.4381 &247 &7  &&&& \\
    1005.6946 &0.5848 &241 &9  &&&& \\
    1006.7059 &0.7352 &82  &11 &&&& \\
    1008.7135 &0.0337 &190 &1  &&&& \\
    1009.7077 &0.1816 &196 &3  &&&& \\
    1010.7002 &0.3291 &144 &5  &&&& \\  \hline
    November 1998 &&&&&&& \\ \hline
    1121.4971 &0.8064 &136 &6&&& \\
    1122.4786 &0.9523 &259 &7&&& \\
    1123.4666 &0.0993 &245 &1&&& \\
    1124.4742 &0.2491 &183 &2&&& \\
    1125.4906 &0.4003 &139 &3&&& \\
    1126.4920 &0.5492 &113 &4&&& \\
    1127.4864 &0.6971 &162 &5&&& \\ \hline
    \end{tabular}
  \end{center}
\end{table*}

\begin{table*}
  \begin{center}
    \caption{Summary of the observations 1999-2002.}
    \label{table3}
    \begin{tabular}{lccclccc}
    \hline
    July-August 1999 &&& &September 1999 &&& \\ \hline
    1383.7017 &0.7998 &248 &9  &1443.4928 &0.6915 &179 &4\\
    1384.7178 &0.9509 &238 &11 &1443.6184 &0.7102 &164 &5\\
    1385.7232 &0.1004 &217 &1  &1444.4788 &0.8382 &230 &6\\
    1386.7124 &0.2475 &229 &3  &1445.5656 &0.9998 &194 &7\\
    1387.7094 &0.3958 &200 &5  &1447.6084 &0.3036 &174 &1\\
    1388.7278 &0.5472 &181 &7  &1448.6212 &0.4542 &184 &2\\
    1389.7031 &0.6923 &178 &8  &1449.5979 &0.5994 &178 &3\\
    1390.7198 &0.8435 &151 &10 &&&&\\
    1391.7045 &0.9899 &243 &12 &&&&\\
    1392.7108 &0.1396 &258 &2  &&&&\\
    1393.7275 &0.2908 &189 &4  &&&&\\
    1394.7386 &0.4411 &206 &6  &&&&\\  \hline
    October 1999 &&&&&&& \\ \hline
    1471.5553 &0.8648 &267 &4&&&\\
    1472.5753 &0.0165 &249 &1&&&\\
    1473.5795 &0.1658 &223 &2&&&\\
    1475.5762 &0.4628 &221 &3&&&\\
    1505.3230 &0.8865 &128 &5&&&\\ \hline
    August 2001      &&& &August 2002 &&&\\ \hline
    2120.5503 &0.3796 &192 &3   &2507.6078 &0.9405 &139 &9 \\
    2121.5642 &0.5303 &225 &4   &2508.6582 &0.0967 &208 &1 \\
    2122.7102 &0.7008 &185 &6   &2509.5877 &0.2349 &200 &3 \\
    2123.6083 &0.8344 &196 &8   &2510.5909 &0.3842 &216 &4 \\
    2124.5644 &0.9765 &193 &10  &2511.5413 &0.5255 &195 &5 \\
    2125.6758 &0.1418 &177 &1   &2512.6316 &0.6876 &212 &7 \\
    2126.7146 &0.2963 &189 &2   &2513.5494 &0.8242 &179 &8 \\
    2128.6067 &0.5777 &166 &5   &2514.5854 &0.9782 &206 &10 \\
    2129.6470 &0.7324 &190 &7   &2515.5768 &0.1257 &207 &2 \\
    2130.6415 &0.8804 &208 &9   &2518.5749 &0.5715 &129 &6 \\ \hline
    November 2002    &&& &&&&\\ \hline
    2588.3964 &0.9549 &207 &7 &&&& \\
    2589.3764 &0.1007 &160 &1 &&&& \\
    2591.3567 &0.3951 &241 &3 &&&& \\
    2592.3705 &0.5459 &219 &4 &&&& \\
    2597.4333 &0.2988 &252 &2 &&&& \\
    2599.4798 &0.6031 &275 &5 &&&& \\
    2600.4896 &0.7532 &259 &6 &&&& \\ \hline
    \end{tabular}
  \end{center}
\end{table*}

\newcommand{\etal}{et al.}

\begin{acknowledgements}
  The results presented in this manuscript are based on observations
  made with the Nordic Optical Telescope, operated on the island of La
  Palma jointly by Denmark, Finland, Iceland, Norway, and Sweden, in
  the Spanish Observatorio del Roque de los Muchachos of the Instituto
  de Astrofisica de Canarias. The authors are grateful for the
  fruitful discussions with Jaan Pelt, Oleg Kochukhov, Thorsten
  Carroll and Markus Kopf.
\end{acknowledgements}

\end{document}